\documentclass[aps,prl,twocolumn,groupedaddress,showpacs]{revtex4}
\usepackage{graphicx}
\usepackage{amsmath}
\usepackage{amsfonts}
\usepackage{graphicx}
\usepackage{hyperref}

\newcommand{\delete}[1]{}

\newcommand{\be}{\begin{equation}}
\newcommand{\ee}{\end{equation}}

\begin{document}

\title{Sensing Short-Range Forces with a Nanosphere Matter-Wave Interferometer}
\author{Andrew Geraci}
\affiliation{Department of Physics, University of Nevada, Reno, NV}
\email[]{ageraci@unr.edu}
\author{Hart Goldman}
\affiliation{Department of Physics, Stanford University, Stanford, CA}
\email[]{hgoldman@stanford.edu}

\begin{abstract}
We describe a method for sensing short range forces using matter wave interference in dielectric nanospheres. When compared with atom interferometers, the larger mass of the nanosphere results in reduced wave packet expansion, enabling investigations of forces nearer to surfaces in a free-fall interferometer. By laser cooling a nanosphere to the ground state of an optical potential and releasing it by turning off the optical trap, acceleration sensing at the $10^{-8}$m/s$^2$ level is possible. The approach can yield improved sensitivity to Yukawa-type deviations from Newtonian gravity at the $5$ $\mu$m length scale by a factor of $10^4$ over current limits.
\end{abstract}


\maketitle

{\it{Introduction.}}
Light-pulse atom interferometers have been demonstrated as a powerful tool for precision sensing, enabling gravimetry at the $10^{-9}g$ level \cite{chu,kasevichg,muller}, gravity gradiometry at the $10^{-9}$ s$^{-2}/\sqrt{\rm{Hz}}$ level \cite{kasevichgrad}, and rotation sensing at the $10^{-8}$ rad/s$/\sqrt{\rm{Hz}}$ level \cite{kasevichrot}. Atom interferometers can also be used in principle for measuring the gravitational attraction of nearby masses \cite{AIbigG}, and for tests of deviations from Newton's inverse square law of gravitation \cite{add,sg,savasandy,islreview,wolf,tino,wacker}. In addition atom interferometers can be used as a surface probe for electromagnetic forces \cite{lev} such as Casimir-Polder forces \cite{casimir,casimirpolder, hinds, cornell}. A challenge for applying light-pulse atom interferometers to such measurements in proximity to surfaces results from the finite wave packet expansion of the atomic cloud. By replacing the atom with a massive dielectric object which is laser-cooled to its motional ground state in an optical trap, the velocity spread dramatically decreases as $(m_{\rm{a}}/M)^{1/2}$, where $m_{\rm{a}}$ and $M$ are the mass of the atom and sphere respectively, enabling measurement times of order $1$ second with a wave packet spread of order $1$ $\mu$m.

In this paper we describe two protocols which utilize macroscopic matter wave phenomena in dielectric spheres to perform sensitive acceleration measurements near material surfaces.  First we describe a near-field Talbot interferometer \cite{talbot,talbot-lau,nimmrichter} which diffracts a sphere from a pure phase grating made of light to generate a density distribution with a fringe pattern at twice the grating period.  Such a setup can be used as an accelerometer to test for corrections to Newtonian gravity at short range. These corrections are generally
parameterized according to a Yukawa-type potential \be
V(r)=-\frac{G_Nm_1m_2}{r}\left[1+\alpha e^{-r/\lambda}\right],
\label{graveq}\ee where $m_1$ and $m_2$ are two masses interacting
at distance $r$, $\alpha$ is the strength of the potential relative
to gravity, and $\lambda$ is the range of the interaction. For two
objects of mass density $\rho$ and linear dimension $\lambda$ with
separation $r \approx \lambda$, a Yukawa-force scales roughly as
$F_Y \sim G_N \rho^2 \alpha \lambda^4$, decreasing rapidly with
smaller $\lambda$.  We estimate sufficient sensitivity to measure $\alpha=400$ at the $\lambda=5$ $\mu$m length scale in such a setup. The current experimental limits at $5$ $\mu$m are $|\alpha| > 3 \times 10^6$ \cite{stanford08}.

 We then compare this to a ballistic experiment which is not based on interference, in which a larger nanosphere is initially cooled to the ground state of an optical potential. After cooling, the optical trap is ramped down quickly allowing the sphere to undergo free wave packet expansion at a rate determined by the ground state momentum spread. For $200$ nm diameter spheres, such an approach has a sensitivity of $1$ $\mu$Gal $=10^{-8}$m/s$^2$, which is comparable to falling corner cube gravimetry systems \cite{faller}. We compare the two techniques as a function of the the temperature and mass of the nanosphere and conclude with a discussion of the systematic error and noise sources for each measurement protocol.




{\it{Protocol}.} A diagram of the interferometry protocol is given in Fig. 1, and a list of experimental parameters is given in Table \ref{paramTable}. We consider a silica sphere of radius $R=6.5$ nm which is optically trapped and cooled such that its center-of-mass wave function $\psi_{CM}$ is near the harmonic oscillator ground state with oscillator frequency $\omega$ determined by the trap. The sphere is then released from the trap and allowed to fall freely in the $z$-direction (with the $x$-direction being transverse to the fall) next to a wall behind which a mass can be placed. 
Immediately after it is released, the wave function has transverse spread $\sigma_{x}=\sqrt{\frac{\hbar}{2 M\omega}}\approx 6$ nm. After one Talbot time $T_T=Md^2/h$, the wave packet has expanded and the sphere is diffracted by a pure phase grating of period $d = 0.25$ $\mu$m. The sphere is then allowed to propagate a time $T_T$ after the grating to a position-sensitive detector, which can be an optical cavity \cite{kippenbergdisp} or split photodetector. Such detectors can achieve sub pm$/\sqrt{\rm{Hz}}$ position resolution, which is adequate for the proposed measurements. After recording the position of the sphere after several experiments, an interference pattern builds up one measurement at a time, with fringes spatially separated by $2d$. An acceleration $a$ in the transverse direction due to the presence of the wall results in a shift in the fringe pattern by an amount $\delta x_\phi=-aT_T^2$. A measurement of the influence of the gravitational attraction of the mass can then be obtained from the relative phase between the fringe patterns with and without the presence of the mass behind the wall.

\begin{figure}[!t]
\begin{center}
\includegraphics[width=0.8\columnwidth]{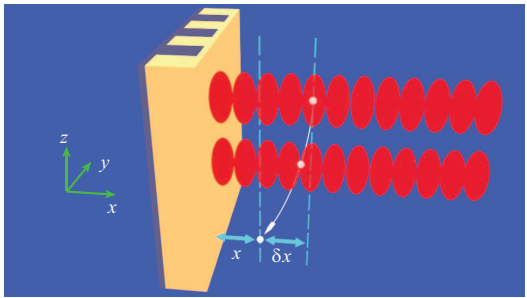}
\caption{Proposed experimental setup. A nanosphere is cooled in an optical trap, and allowed to fall for $T_T$ after which a light pulse grating is applied. After another time $T_T$ the position of the center of mass of the sphere is recorded. Such measurements combine to reveal an interference pattern, where the node positions depend on the transverse $(x)$ acceleration experienced by the bead throughout its fall. The centroid of the distribution also shifts towards the wall by an amount $\delta x$ as a result of the acceleration. The wall consists of vertical sections of varying density to modulate the gravitational acceleration depending on the initial $y$-position of the trap.
\label{setup}}
\end{center}
\end{figure}

\begin{table}
\centering
\begin{tabular}{l | c | c}

\hline
\hline
Parameter & Symbol & Value \\
\hline
sphere radius & $R$ & $6.5$ nm \\
sphere density & $\rho$ & 2300 $\text{kg}/\text{m}^3$ \\
dielectric constant & $\epsilon$ & 2 \\
trap frequency & $\omega$ & $2 \pi \times 100$ Hz \\
grating period & $d$ & $0.25$ $\mu$m \\
grating peak intensity & $I$ & $55$ kW/m$^2$ \\
\hline
\hline
\end{tabular}
\caption{Experimental parameters for the interference protocol corresponding to a total fall time of $2T_T=0.5$s.}
\label{paramTable}
\end{table}

The effect of the grating on the wave function can be understood using the phase-space formalism of \cite{TimeDomainIonizing}. We utilize the transverse Wigner distribution associated with the sphere's center of mass (CM) mode
\begin{equation}
w_0(x,p)=\frac{1}{2\pi\hbar}\int_{-\infty}^{\infty} ds e^{isp/\hbar}\langle x-s/2|\hat{\rho}|x+s/2\rangle
\end{equation}
where $\hat{\rho}$ is the density matrix for the CM mode. If the sphere is cooled to the ground state of center of mass motion, 
$w_0(x,p)=A\cdot\exp\left[-\frac{x^2}{\sigma_x^2}-\frac{p^2}{\sigma_p^2}\right]$ where $A$ is fixed by normalization. If there is any transverse acceleration $a$, after falling for a time $t$, the Wigner distribution is sheared accordingly as $w_1(x,p;t)=w_0(x-\frac{p}{M}t+\frac{1}{2}at^2,p-Mat)$, where $w_0$ is the Wigner distribution at time $t=0$. At the grating, the wave function undergoes a transformation of the form $|\psi_{CM}\rangle\mapsto U|\psi_{CM}\rangle$, so the density matrix transforms as $\rho\mapsto U\rho U^\dagger$. Since the de Broglie wavelength of the sphere is very small, we can employ the eikonal approximation $U(x)=\exp(i\phi(x))$, where $\phi(x)=-\frac{1}{\hbar}\int_{-\infty}^{\infty}V(x,t)dt=\frac{\alpha_\omega I \tau}{\hbar c \epsilon_0}\sin^2(\pi x/d)
\equiv \phi_0\sin^2(\pi x/d)$ where $\alpha_\omega=4\pi R^3\epsilon_0(\frac{\epsilon-1}{\epsilon+2})$ is the polarizability of the sphere, $I$ is the peak laser intensity, and $\tau=1$ $\mu$s is the pulse duration. After propagating to the grating over a time $t_0$, the sheared Wigner distribution transforms via the integration kernel
\begin{equation}
w_2(x,p; t_0)=\int_{-\infty}^{\infty}dp_0dx_0K(x,p;x_0,p_0)w_1(x_0,p_0; t_0)
\end{equation}
where
\begin{widetext}
\begin{eqnarray}
K(x,p;x_0,p_0)&=&\frac{1}{2\pi\hbar}\int dsds_0 e^{i(p_0s_0+ps)/\hbar}\langle x-s/2|U|x_0+s_0/2\rangle\langle x+s/2|U^*|x_0-s_0/2\rangle \\
&=&\frac{1}{2\pi\hbar}\delta(x-x_0)\sum_{j,m\in\mathbb{Z}}b_jb_{j-m}^*e^{2\pi i mx/d}\delta\left(p-p_0-(j-m/2)\frac{2\pi\hbar}{d}\right)
\label{eq: kernel}
\end{eqnarray}
\end{widetext}
The $b_m=(-i)^me^{i\phi_0/2}J_m(\phi_0/2)$ are called Talbot-Lau coefficients \cite{KDTL}. 

\begin{figure}[!t]
\begin{center}
\includegraphics[width=0.8\columnwidth]{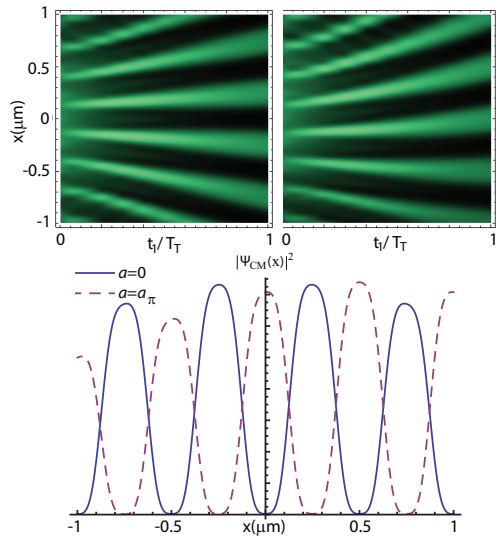}
\caption{(left) Density plot of $|\Psi(x)|^2$ following the grating for zero acceleration for releasing the trap at $\omega_0=2\pi \times 100$ Hz from its ground state.  (right) As in left panel, with $a_\pi=4 \times 10^{-7} g$ constant acceleration. (lower) Lineouts taken at $t_1=T_T$ after the grating for $a=0$ (solid) and $a=a_\pi$ (dashed).
\label{fringes}}
\end{center}
\end{figure}

After propagating for an additional time $t_1$ to the detector, the final fringe pattern can be obtained by integrating over momentum
\begin{eqnarray*}
W_3(x)&=&|\psi_{CM}(x)|^2\\
&=&\int_{-\infty}^\infty dp w_2(x-\frac{p}{M}t_1+\frac{1}{2}at_1^2, p-Mat_1)
\end{eqnarray*}
We find an optimum $\phi_0 \approx 1.5$. For the parameters given in Table \ref{paramTable},
this corresponds to an intensity of $I=55$ kW/${\rm{m}}^2$.
The final fringe pattern develops a phase $\Phi(a,t_0,t_1)$ which is proportional to the transverse acceleration $a$ to lowest order
\begin{equation}
\Phi(a,t_0,t_1)\approx\left(\frac{i\pi t_0t_1}{d}\right) a
\end{equation}
For $t_0=t_1=T_T$ and $R=6.5$ nm, the transverse acceleration required for a $\pi$ phase shift in the expected fringe pattern is approximately $a_\pi=d/T^2_T=4$ $\mu\text{m}/\text{s}^2=4 \times 10^{-7}g$. This means that the acceleration sensitivity of the experiment is fully determined by the grating period and time of fall. If the sphere had an initial transverse momentum kick, the phase shift $\Phi$ remains the same, which means that while this experiment is highly sensitive to transverse acceleration, it is insensitive to any systematic initial momentum kicks.  Plots of the probability densities we obtain for parameters given in Table \ref{paramTable} are shown in Fig. (\ref{fringes}) for $a=0$ and $a=a_\pi$.

The above assumes that sphere can be cooled to its ground state of CM motion. However, it is possible to obtain a fringe pattern at temperatures above that of the ground state, where the CM wave function becomes a superposition of harmonic oscillator eigenstates. The effect of temperature on the Wigner distribution can be approximated as a widening of the position and momentum spreads of the pure ground state distribution 
according to $k_BT \approx\hbar\omega\bar{n}(T)\approx M\omega^2\sigma_x^2(T)$, where $\bar{n}(T)$ is the average principal quantum number of the CM state at temperature $T$. Therefore the position (and momentum) spread grow as $\sqrt{\bar{n}(T)}$ at large $T$.

{\it{Short-range force measurements.}}
The Casimir-Polder force between a small dielectric sphere
and metal plane can be written as
\cite{casimirpolder} $F_{\rm{cp}}=-\frac{3\hbar c
\alpha_\omega}{8\pi^2\epsilon_0} \frac{1}{z^5}$.  This force results in an acceleration of $4 \times 10^{-7}g$ on the sphere and displaces the fringe pattern by approximately $\pi$ for a surface separation of $10$ $\mu$m and $R=6.5$ nm, $T_T=0.25$s. The phase shift is $\approx 3\pi$ for $R=5$ nm, $T_T = 0.1$ s, and surface separation of $6$ $\mu$m. Thus by averaging over $10^4$ shots, the Casimir-Polder acceleration can be measured at or below the percent level.

\begin{figure}[!t]
\begin{center}
\includegraphics[width=1.0\columnwidth]{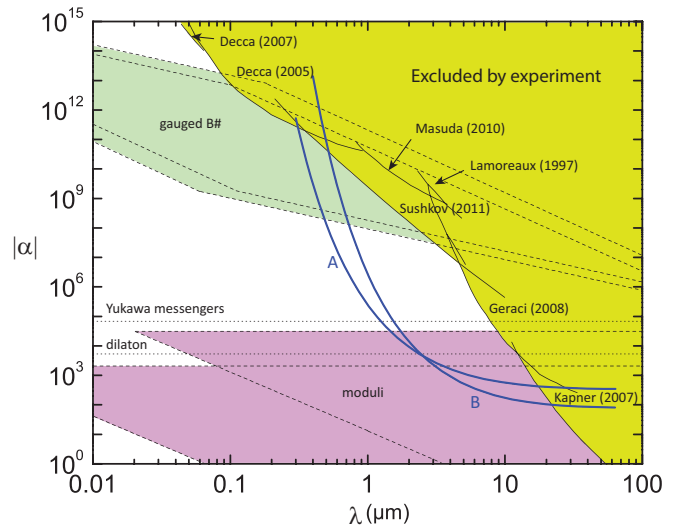}
\caption{Current experimental bounds \cite{stanford03,stanford08,kapner,decca1,decca2,masuda,lamoreaux1,sushkov} and theoretical predictions \cite{savasandy} for a non-Newtonian potential of the form $V(r)=-\frac{G_N m_1 m_2}{r}\left[1+\alpha\exp(-\frac{r}{\lambda})\right]$ between two masses $m_1$ and $m_2$ separated by $r$. Curves A and B are the predicted sensitivities for wall-separation $6$ $\mu$m and $10$ $\mu$m, with corresponding Talbot times of $0.1$ s and $0.25$ s, respectively.
\label{alphalambda}}
\end{center}
\end{figure}

For a short-distance gravity measurement, we consider the differential shift in the fringe pattern between the case where the sphere falls next to a gold section of the wall and the case where the sphere falls next to a silicon section of the wall.  Here the shift from the Casimir acceleration is common to both cases as a uniform gold coating covers the surface of the wall. We take the width of the gold and silicon sections to be $40$ $\mu$m. We consider two cases, with $R=6.5$ nm, $T_T=0.25$ s and a $10$ $\mu$m separation of the sphere from the wall, and with $R=5$ nm, $T_T=0.1$ s and $6$ $\mu$m sphere-wall separation.  Projected sensitivity is shown in Fig. \ref{alphalambda} for a phase resolution of $\pi/300$, corresponding to averaging over $10^5$ shots of the experiment.

\begin{figure}[!t]
\begin{center}
\includegraphics[width=0.9\columnwidth]{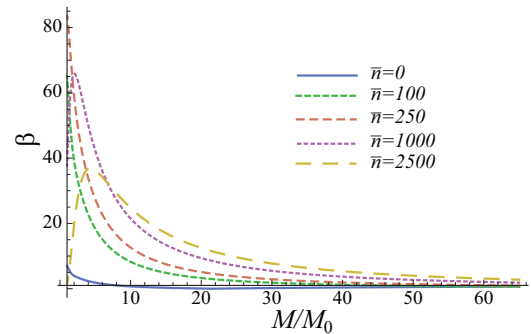}
\caption{The improvement factor $\beta$ for an interference experiment relative to a ballistic wave-packet expansion experiment as a function of mass at various temperatures. The fall time is fixed at $t=0.5$ s, corresponding to  $2 T_T$ for a sphere of mass $M=M_0$, where $M_0 \equiv \frac{4}{3}\rho\pi R^3$ with $R=6.5$ nm. 
\label{winplot}}
\end{center}
\end{figure}

{\it{Comparison of Interference and Ballistic Measurements.}}
It is interesting to compare the position sensitivity for a ballistic approach where the position of the sphere is measured after falling from the trap versus the Talbot interferometer sensitivity.  Assuming the particle is cooled to the ground state in the harmonic trap, the velocity spread due to zero point motion $\sigma_v = \sqrt{\frac{\hbar \omega}{2 M}}$ will cause a spread in the measured position of the bead after it falls during the experiment. The position spread after a measurement at time $t$ later is thus given by $\sigma_v t$.  After $N$ repeated experiments, the uncertainty in the mean of the distribution goes as $\sigma_v t / \sqrt{N}$.  This uncertainty is added to the signal $\delta x = \frac{1}{2} a t^2$ for an acceleration $a$ towards the wall.  In the Talbot interferometer, the fringe pattern shifts by a comparable amount due to the acceleration from the wall.  However, the momentum uncertainty in the ground state harmonic oscillator trap does not influence the location of particular interference fringes $-$ only 
 the overall envelope is influenced by the initial momentum spread. The uncertainty in the fringe position of the fringe maxima when taking $N$ measurements is $\sim d/\sqrt{N}$.  Since the period is known, the fringe pattern can be fit using a function with a known period and variable phase.  The improvement over the ballistic measurement is given by $\beta=\chi \sigma_v t/d$, where $\chi$ is the fringe contrast of the interferometer. Plots of $\beta$ as a function of mass are given for various temperatures in Fig. \ref{winplot} for a fixed fall time of $t=2T_T=0.5$ s.

At temperatures near the ground state temperature, increasing mass results in increasing localization and ultimately a wave packet which is too narrow to interact with the grating, at which point the interference signal vanishes. For higher temperatures, the position and velocity spreads become large enough to let higher masses interact with the grating, but velocity spreads which are large compared to the grating velocity $v_g=\hbar/dM$ result in reduced contrast. Increases in the mass actually improve the sensitivity of the interference setup until one strays too far from the Talbot condition (i.e. $T_T(M) >> t$). At this point the contrast of the fringe pattern falls until there is no more visibility. Thus, at higher temperatures, $\beta$ increases for some mass interval, peaks, and subsequently falls to zero.

If the mass is increased without also extending the time-of-flight to maintain the Talbot time condition, the ballistic experiment eventually exceeds the sensitivity to the interference experiment, with the caveat that the location of the sphere at the end of the experiment is sensitive to systematic errors in the initial velocity distribution upon release from the optical trap.  For a 100 nm radius sphere initially cooled to the ground state and allowed to expand and fall ballistically, the sensitivity curves are approximately $10 \times$ better than those shown in Fig. \ref{alphalambda}, however with this additional source of error.

\emph{Systematics}.
While the effect of decoherence due to gas and blackbody radiation is negligible for our setup, Rayleigh scattering of photons from the laser grating can result in decoherence in the interference experiment. The timescale for this however can be much larger than the time of interaction with the grating, which we take to be $1\mu$s. 
Since the spread of the CM wave function at the grating is of order $\sim d$, 
the decoherence time is roughly the time for one scattering event, approximately $1$ ms. 

A deviation in vertical alignment will produce a constant offset in the measured acceleration. If each shot has a varying misalignment this becomes an additional noise source. Such noise is negligible for an angular stability of $\sim 0.5$ ppm.

While the fringe locations are insensitive to any systematic velocity kick given to the falling sphere as it is released from the optical standing wave trap, the setup is sensitive to vibrational noise in the mirrors during the application of the grating pulse and the during detection of the sphere. Maximal sensitivity requires vibrational stability of $\sim 10^{-3}$ $\mu$m$/\sqrt{\rm{Hz}}$ at frequencies around $1$ Hz.


The polycrystalline structure of the gold coating on the wall results in local electric variations due to the patch effect \cite{patch,patch2}. These patch potentials can drift with time and vary over spatial extent of the wall. We can estimate the acceleration applied to the sphere as a result of typical patch potentials ranging from $\sim 50$ mV variations over length scales of a few $\mu$m to be of order $10^{-7}g$. Such accelerations will contribute to the fringe shift of the interferometer and would need to be characterized experimentally. However, the initial trap can be translated laterally, so that the experiment can first be performed with the sphere closest to a gold section of the wall, and then closest to the adjacent silicon section of the wall, and then closest to the next gold section etc. as shown in Fig. \ref{setup}. Thus, by scanning the initial $y$-position of the sphere along the wall, one expects a spatially periodic signal for the acceleration due to the mass. The variation of the acceleration due to the patch effect is not expected to exhibit the same periodicity as the underlying spatial density pattern in the wall. This can be used in principle to distinguish the effects.

{\it{Discussion.}}
 The matter wave accelerometer we have presented can be advantageous when compared with light-pulse atom interferometry for use in surface-force measurements where localization of the sensor is required. This technique could lead to advances in tests of inverse-square law violations of gravity at $\mu$m distances and Casimir-Polder force measurements in new regimes.  

{\it{Acknowledgements.}}
We thank A. Arvanitaki who contributed greatly to the development of this work. We thank S. Dimopoulos, J. Weinstein, and G. Ranjit for discussions. AG is supported in part by grant NSF-PHY 1205994.

\end{document}